\documentclass[12pt]{iopart}
\usepackage{iopams}  

\usepackage{epsf}
\usepackage{rotating}

\usepackage{graphicx,epsfig}
\usepackage{dcolumn}
\usepackage{bm}

\def\cA{{\cal A}}

\newcommand{\req}[1]{Eq.~(\ref{#1})}
\newcommand{\avg}[1]{\langle #1\rangle}

\begin{document}

\title[Optimal Resource Allocation with Transportation
Bandwidths]{Optimal Resource Allocation in Random Networks 
with Transportation Bandwidths}

\author{C.~H.~Yeung and K.~Y.~Michael Wong}

\address{Department of Physics, The Hong Kong University
of Science and Technology, Hong Kong, China}
\eads{\mailto{phbill@ust.hk}, \mailto{phkywong@ust.hk}}
\begin{abstract}
We apply statistical physics to study the task of resource allocation
in random sparse networks with limited bandwidths for the 
transportation of resources along the links.
Recursive relations from the Bethe approximation are converted into useful algorithms.
{\it Bottlenecks} emerge when the bandwidths are small,
causing an increase in the fraction of idle links.
For a given total bandwidth per node,
the efficiency of allocation increases with the network connectivity.
In the high connectivity limit, 
we find a phase transition at a critical bandwidth,
above which clusters of balanced nodes appear,
characterised by a profile of homogenized resource allocation 
similar to the Maxwell's construction.

\end{abstract}

\maketitle

\section{Introduction}
\label{sec:Introduction}

Analytical techniques developed in the statistical physics of 
spin models have been widely employed in the analysis of systems 
in a wide variety of fields,
such as neural networks \cite{hertz, nishimori},
econophysical models \cite{challet}, 
and error-correcting codes \cite{nishimori, kabashima}.
Analogy is drawn between the interactions and dynamics commonly
present in the spin models and other systems.
Consequently, 
notions such as cost functions and noises can find their corresponding physical
quantities in spin models.
This parallelism contributes to the success of statistical mechanics in various fields.

Recently,
a statistical physics perspective was successfully applied to the problem
of resource allocation on sparse random networks \cite{wong2006, wong2007}.
Resource allocation is a well known network problem in the areas of computer science
and operations management \cite{peterson, ho}.
It is relevant to applications such as load balancing in computer networks,
reducing Internet traffic congestion, and streamlining network flow
of commodities \cite{shenker, rardin}.
In a typical setup,
each node of the network has its own demand or supply of resources,
and the task is to transport the resources through the links to 
satisfy the demands while a global transportation cost function is optimized.

The work in \cite{wong2006, wong2007} can be extended to consider the effects of 
bandwidths of the transportation links.
In communication networks,
connections usually have assigned bandwidths.
Bandwidths limit the currents flowing in the links or, 
in equivalent models, the interaction strengths defined on the links 
connecting the site variables.
The significance of finite bandwidths was recently recognized in several 
similar problems.
For example, transportation problems with global constraints
on interaction strengths were studied, 
through a model of transportation network
with limited total conductance \cite{bohn},
and another model of transportation network with 
limited total pipe volume or surface area which increase
the resistance of pipes \cite{durand}.
Among these studies, constraints on interaction strengths were
implemented as limits on conductance.

The purpose of this paper is twofold.
First,
we demonstrate the close relation between statistical mechanics and 
distributed algorithms.
Such close relations have facilitated statistical mechanics to provide insights
to a number of principled algorithms recently,
as illustrated by the relation between Bethe approximation
and error-correcting codes \cite{nishimori, opper1999, vicente1999},
and probabilistic inference \cite{mackay2003},
as well as the replica symmetry-breaking ansatz and the survey propagation
algorithm in $K$-satisfiability problems \cite{mezard2002}.
These contributions were made in problems with discrete variables.
In this paper,
we extend the Bethe approximation to the resource allocation involving continuous variable
with finite bandwidths.
Since bandwidths limit the currents along the links,
both the message-passing and price iteration algorithms 
proposed in \cite{wong2006, wong2007} have to be modified,
which enable us to find the optimal solutions
without the need of a global optimizer.

The second purpose is to study the behavior of the optimized networks
when the connectivity and bandwidth changes,
using insights generated from the recursion relations of the free energy.
We observe a number of interesting physical phenomena in networks with finite bandwidths.
For example,
there is the emergence of {\it bottlenecks} when the bandwidth decreases,
resulting in the shrinking of the fraction of unsaturated links.
We also identify the correlations between the 
capacities of the nodes and their roles as sources,
sinks and relays.
We find that resources are more efficiently distributed with increasing
connectivity.
Scaling relations are found in the distributions of currents and chemical potentials.
In the high connectivity limit, 
we find a phase transition at a critical bandwidth,
above which clusters of balanced nodes appear,
characterised by a profile of homogenized resource distribution
reminiscent of the Maxwell's construction in thermodynamics.
When adapted to scale-free networks,
changes in this profile enable us to identify the enhanced homogeneity of resources
brought by the presence of hubs to nodes with low connectivity.

The paper is organized as follows. 
In Section \ref{sec:model}, 
we introduce the general model,
and the analysis is presented in Section \ref{sec:analysis}.
We demonstrate the conversion of the derived recursive equations
into algorithms in Section \ref{sec:MP}.
In Section \ref{sec:sim}, 
we compare the analytic solutions with numerical simulations in networks
with low connectivity,
and report on the bottleneck effect.
We will examine the limit of high connectivity in Section \ref{sec:highC}.
We conclude the paper and point out the potential applications of the model
in Section \ref{sec:Conclusion}.

\section{The model}
\label{sec:model}

We address the problem of resource allocation on random sparse networks of
nodes and links.
Each node of the network has its own supply or demand of resources,
and the task is to transport the resources through the links to satisfy
the demands while optimizing a cost function of transportation costs.
The amount of resources transported through a link is limited by the bandwidth,
which leads to extra constraints in the resource allocation problem.

Bandwidth limitations require us to generalize the problem in the following way.
In the case of infinite bandwidths,
currents along the links can be set arbitrarily large to satisfy the demands of all nodes.
Hence hard constraints on the satisfiability of the node demands can be realized,
provided that the networkwide supply of resource is greater than their
resource demand.
On the other hand,
in the present model of finite bandwidth,
a node with resource demand can still experience shortage even though its 
neighbors have adequate supply of resources, 
since the provision of resources can be limited by the bandwidths of the links.
Hence,
the hard constraints on node satisfaction of the infinite-bandwidth model is 
replaced by soft constraints,
and the energy function is generalized to include the cost of unsatisfaction,
or shortage of resources.

Specifically,
we consider a network with $N$ nodes, 
labelled $i\!=\!1,\dots,N$.
Each node $i$ is randomly connected to $c$ other nodes.
The connectivity matrix is given by $\cA_{ij}=1, 0$ for connected and unconnected
node pairs respectively.
Except for the discussion on scale-free networks in
Section \ref{sec:highC} B4,
we focus on sparse networks,
namely, 
those of intensive connectivity $c\sim O(1)\ll N$.

Each node $i$ has a capacity $\Lambda_i$ randomly drawn from a distribution
$\rho(\Lambda_i)$.
Positive and negative values of $\Lambda_i$ correspond to supply and demand
of resources respectively.
The task of resource allocation involves transporting resources between
nodes such that the demands of the nodes can be
satisfied to the largest extent.
Hence we assign $y_{ij}\equiv-y_{ji}$ to be the {\it current} drawn from node $j$ to $i$,
aiming at reducing the {\it shortage} $\xi_i$ of node $i$ defined by
\begin{eqnarray}
\label{xi_define}
	\xi_i=\max\biggl(-\Lambda_i-\sum_{(ij)}\cA_{ij}y_{ij}, 0\biggr).
\end{eqnarray}
The magnitudes of the currents are bounded by the {\it bandwidth} $W$,
i.e., $|y_{ij}|\leq W$.
For simplicity and clarity,
we deal with the case of homogeneous connectivity and bandwidth in the paper,
but the analyses and algorithms can be trivially generalized to handle real networks
having heterogeneous connectivity and bandwidth.

To minimize the shortage of resources after their allocation,
we include in the total cost both the shortage cost and the transportation cost.
Hence,
 the general cost function of the system can be written as 
\begin{eqnarray}
\label{E_define}
	&&E=R\sum_{(ij)}\cA_{ij}\phi(y_{ij})
	+\sum_i\psi(\Lambda_i,\{y_{ij}|\cA_{ij}=1\}).
\end{eqnarray}
The summation $(ij)$ corresponds to summation over all node pairs,
and $\Lambda_i$ is a quenched variable defined on node $i$.

In the present model of resource allocation, 
the first and second terms correspond to the transportation and shortage 
costs respectively.
The parameter $R$ corresponds to the {\it resistance} on the currents,
and $\Lambda_i$ is the capacity of node $i$.
The transportation cost $\phi(y_{ij})$ can be a general even function
of $y_{ij}$.
(Our model can be generalized to consider transportation costs $\phi$
that are odd functions of $y_{ij}$.
In such cases,
one simply has to introduce extra quenched variables of the 
links representing the favored directions of the currents.)
In this paper, 
we consider $\phi$ and $\psi$ to be concave functions of their arguments,
that is,
$\phi'(y)$ and $\psi'(\xi)$ are non-decreasing functions.
Specifically, 
we have the quadratic transportation cost $\phi(y)=y^2/2$, 
and the quadratic shortage cost $\psi(\Lambda_i, \{y_{ij}|\cA_{ij}=1\})=\xi_i^2/2$.
As considered previously,
other concave and continuous functions of $\phi$,
such as the anharmonic function in \cite{wong2007},
result in similar network behavior.

The model can also be extended to probabilistic inference on graphical models \cite{jordan}.
In this context $y_{ij}=y_{ji}$ may represent the coupling between observables
in nodes $j$ and $i$,
$\phi(y_{ij})$ may correspond to the logarithm of the prior distribution 
of $y_{ij}$,
and $\psi(\Lambda_i, \{y_{ij}|\cA_{ij}=1\})$ the logarithm of the likelihood
of the observables $\Lambda_i$.

\section{Analysis}
\label{sec:analysis}

The analysis of the model is made convenient by the introduction 
of the variables $\xi_i$.
It can be written as the minimization of \req{E_define} in the space of 
$y_{ij}$ and $\xi_i$,
subject to the constraints
\begin{eqnarray}
\label{xiCon}
	&&\Lambda_i+\sum_{(ij)}\cA_{ij}y_{ij}+\xi_i\ge 0 ,
	\nonumber\\
	&&\xi_i\ge 0 ,
\end{eqnarray}
and the constraints on the bandwidths of the links
\begin{eqnarray}
	W-y_{ij}\ge 0 ,
	\nonumber\\
	W+y_{ij}\ge 0.
\end{eqnarray}

We consider the dual of the original optimization problem.
Introducing Lagrange multipliers to the above inequality constraints,
the function to be minimized becomes 
\begin{eqnarray}
\label{Lagr}
	L\!=\!\sum_i\biggl[\psi(\xi_i)+\mu_i\biggl(\Lambda_i+\sum_{(ij)}\cA_{ij}y_{ij}+\xi_i\biggr)
	+\alpha_i\xi_i\biggr]
	\nonumber\\
	+\sum_{(ij)}\cA_{ij}\biggl[R\phi(y_{ij})+\gamma^+_{ij}(W-y_{ij})+\gamma^-_{ij}(W+y_{ij})\biggr],
\end{eqnarray}
with the Kuhn-Tucker condition
\begin{eqnarray}
\label{constraints}
	\mu_i\biggl(\Lambda_i+\sum_{(ij)}\cA_{ij}y_{ij}+\xi_i\biggr)=0,
	\nonumber\\
	\alpha_i\xi_i=0,
	\nonumber\\
	\gamma^+_{ij}(W-y_{ij})=0,
	\nonumber\\
	\gamma^-_{ij}(W+y_{ij})=0,
\end{eqnarray}
and the constraints $\mu_i\leq 0, \alpha_i\leq 0$, $\gamma_{ij}^+\leq 0$ and 
$\gamma_{ij}^-\leq 0$.
Optimizing $L$ with respect to $y_{ij}$, 
one obtains
\begin{eqnarray}
\label{solution}
	y_{ij} =Y(\mu_j-\mu_i)
\end{eqnarray}
\begin{eqnarray}
\label{solutionY}
	Y(x) = \max\biggl\{-W, \min\biggl[W, [\phi']^{-1}\biggl(\frac{x}{R}\biggr)\biggr]\biggr\}.
\end{eqnarray}
The Lagrange multiplier $\mu_i$ is referred to as the chemical potential
of node $i$,
and $\phi'$ is the derivative of $\phi$ with respect to its argument.
The function $Y(\mu_j-\mu_i)$ relates the potential difference between
nodes $i$ and $j$ to the current driven from node $j$ to $i$.
For the quadratic cost,
it consists of a linear segment between $\mu_j-\mu_i = \pm WR$ reminiscent of Ohm's law
in electric circuits.
Beyond this range,
$y$ is bounded above and below by $\pm W$ respectively.
Thus,
obtaining the optimized configuration of currents $y_{ij}$ among the nodes 
is equivalent to finding the corresponding set of chemical potentials $\mu_i$,
from which the optimized $y_{ij}$'s are then derived from $Y(\mu_j-\mu_i)$.
This implies that we can consider the original optimization problem
in the space of chemical potentials.

The network behavior depends on whether $\psi'(0)$ is zero or not;
$\psi'(0)$ can be considered as the friction necessary for the chemical potentials 
to overcome when the shortage changes from zero to nonzero.
For the frictionless case of $\psi'(0)=0$,
to which the quadratic $\psi$ considered in this paper belongs,
two types of nodes can be identified. 
Those nodes $i$ with excess resources $(\Lambda_i+\sum_{ij}\cA_{ij}y_{ij}>0)$
are characterized by $\mu_i=0$ and $\xi_i=0$.
Those nodes $i$ with resource shortage are characterized by $\mu_i<0$ and $\xi_i>0$.

For the friction case of $\psi'(0)>0$, 
there is a third type of nodes $i$ with fully utilized resources 
$(\Lambda_i+\sum_{ij}\cA_{ij}y_{ij}=0)$ and characterized by $\mu_i<0$ and $\xi_i=0$.

Three types of links can be identified.
Those links $(ij)$ with $|y_{ij}|=W$ are referred to as the {\it saturated} links.
Those with $0<|y_{ij}|<W$ and $|y_{ij}|=0$ are referred to as {\it unsaturated}
and {\it idle} links respectively.

We introduce the free energy at a temperature $T\equiv\beta^{-1}$,
\begin{eqnarray}
	F=-T\ln Z,
\end{eqnarray}
where $Z$ is the partition function
\begin{eqnarray}
\label{define_Z}
	Z=\prod\limits_{(ij)}\int_{-W}^{W}dy_{ij}\exp\biggl[-\beta\sum_{(ij)}\cA_{ij}\phi{(y_{ij})}
	-\beta\sum_i\psi(\Lambda_i,\{y_{ij}|\cA_{ij}=1\})\biggr].
	\nonumber\\
\end{eqnarray}
The statistical mechanical analysis of the free energy can be carried out using the replica
method or the Bethe approximation,
both yielding the same results \cite{wong2007}.
The replica approach for networks with finite bandwidths is a direct 
generalization of the case with infinite bandwidth in \cite{wong2007}.
Here,
we describe the Bethe approach,
whose physical interpretation is more transparent.

In large sparse networks,
the probability of finding loops of finite lengths on the network is low,
and the local environment of a node resembles a tree.
In the Bethe approximation,
a node is connected to $c$ branches of the tree,
and the correlations among the branches are neglected. 
In each branch, 
nodes are arranged in generations,
A node is connected to an ancestor node of the previous generation,
and another $c-1$ descendent nodes of the next generation.

We consider the vertex $V({\mathbf T})$ of a tree ${\mathbf T}$.
We let $F(y|{\mathbf T})$ be the free energy of the tree when a current
$y$ is drawn from the vertex by its ancestor node.
One can express $F(y|{\mathbf T})$ in terms of the free energies
$F(y_k|{\mathbf T}_k)$ of its descendents $k=1,\dots, c-1$, 
\begin{eqnarray}
\label{betheF}
        &&F(y|{\mathbf T})=-T\ln\Biggl\{
        \prod_{k=1}^{c-1}\left(\int_{-W}^W dy_k\right)
        \exp\biggl[
        -\beta\sum_{k=1}^{c-1}F(y_k|{\mathbf T}_k)       
        \nonumber\\
        &&-\beta R\sum_{k=1}^{c-1}\phi(y_k)
        -\beta\psi\biggl(\max(-\Lambda_{V({\mathbf T})}-\sum_{k=1}^{c-1}y_k+y, 0)\biggr)
        \biggr]\Biggr\},
\end{eqnarray}
where ${\mathbf T}_k$ represents the tree terminated at the $k^{\rm th}$ 
descendent of the vertex, 
and $\Lambda_{V({\mathbf T})}$ is the capacity of $V({\mathbf T})$.
We then consider the free energy as the sum of two parts,
\begin{equation}
\label{eq:free}
	F(y|{\mathbf T})\!=\!N_{\mathbf T}F_{\rm av}\!+\!F_V(y|{\mathbf T}),
\end{equation}
where $N_{\mathbf T}$ is the number of nodes in the tree ${\mathbf T}$,
and $F_{\rm av}$ is the vertex free energy per node.
$F_V(y|{\mathbf T})$ is referred to as the {\it vertex free energy}.
Note that when a vertex is added to a tree,
there is a change in the free energy due to the added vertex.
In the language of the cavity method \cite{mezard1987},
the vertex free energies are equivalent to the {\it cavity fields},
since they describe the state of the system when the ancestor node is absent.
From \req{eq:free}, we obtain a recursion relation for the vertex free energy.
In the zero temperature limit,
this relation reduces to
\begin{eqnarray}
\label{recur}
	&&F_V(y|{\mathbf T})=\min_{\{y_k||y_k|\leq W\}}
	\Biggl[\sum_{k=1}^{c-1}\Biggl(F_V(y_k|{\mathbf T}_k)+R\phi(y_k)\Biggr)
	\nonumber\\
	&&+\psi\Biggl(\max(-\Lambda_V({\mathbf T})-\sum_{k=1}^{c-1}y_k + y, 0)\Biggr)\Biggr] - F_{\rm av}.
\end{eqnarray}
Note that $F_{\rm av}$ has to be subtracted from the above relation,
since the number of nodes increases by 1 when a new vertex is added to the trees.

The average free energy is obtained by considering the average
change in the free energy when a vertex is connected to its $c$ neighbors,
resulting in 
\begin{eqnarray}
	&&F_{\rm av}(y|{\mathbf T})=\Biggl\langle\min_{\{y_k||y_k|\leq W\}}
	\Biggl[\sum_{k=1}^{c}\Biggl(F_V(y_k|{\mathbf T}_k)+R\phi(y_k)\Biggr)
	\nonumber\\
	&&+\psi\Biggl(\max(-\Lambda_V({\mathbf T})-\sum_{k=1}^{c}y_k, 0)\Biggr)\Biggr]\Biggr\rangle_\Lambda.
\end{eqnarray}
where $\Lambda_V$ is the capacity of the vertex $V$ fed by 
$c$ trees ${\mathbf T}_{1}, \ldots, {\mathbf T}_{c}$, and $\langle\dots\rangle_\Lambda$
represents the average over the distribution $\rho(\Lambda)$.
The shortage distribution and current distribution can be 
derived accordingly \cite{wong2007}.

\section{The Message-Passing Algorithm}
\label{sec:MP}

Due to the local nature of the recursion relation (\ref{recur}), 
the optimization problem can be solved by 
message-passing approaches,
which have been successful in problems such as error-correcting cods
\cite{opper1999} and probabilistic inference \cite{mackay2003}.
However, 
in contrast to other message-passing algorithms which pass conditional probability
estimates of discrete variables to neighboring nodes,
the messages in the present context are more complex,
since they are free energy functions $F_V(y|{\mathbf T})$ of the continuous variable $y$.
Inspired by the success of replacing the function messages by their
first and second derivatives in \cite{wong2007},
we follow the same route and simplify the messages devised for this problem.
Note that the validity of this simplification is not obvious in this problem,
owing to the non-quadratic nature of the energy function as a result of the finite
bandwidths.
These additional constraints increase the complexity
of the problem and complicate the messages.

We form two-parameter messages using the first and second derivatives of the vertex free energies.
Let $(A_{ij},B_{ij})\equiv (\partial F_V(y_{ij}|{\mathbf T}_j)/\partial y_{ij},
\partial^2 F_V(y_{ij}|{\mathbf T}_j)/\partial y_{ij}^2)$
be the messages passed from node $j$ to its ancestor node $i$,
based on the messages received from its descendents in the tree ${\mathbf T}_j$.
The recursion relation of the message can be obtained by minimizing the vertex free energy
in the space of the current adjustments $\epsilon_{jk}$ drawn from the descendents.
We minimize 
\begin{eqnarray}
\label{MPL}
       F_{ij}=\sum_{k\ne i}\cA_{jk}\biggl[A_{jk}\varepsilon_{jk}
        +\frac{1}{2}B_{jk}\varepsilon_{jk}^2
        +R\phi'_{jk}\varepsilon_{jk}
        +\frac{R}{2}\phi''_{jk}\varepsilon_{jk}^2\biggr]
        +\psi(\xi_j),
\end{eqnarray}
subject to the constraints
\begin{eqnarray}
\label{MPconstraint}
        \sum_{k\ne i}\cA_{jk}(y_{jk}+\varepsilon_{jk})-y_{ij}+\Lambda_j+\xi_j\ge 0,
\end{eqnarray}
\begin{eqnarray}
        \xi_j \ge 0,
        \nonumber
\end{eqnarray}
together with the constraints on bandwidths, 
\begin{eqnarray}
	|y_{jk}+\epsilon_{jk}|\leq W.
\end{eqnarray}
$\phi_{jk}'$ and $\phi_{jk}''$ represent the first and second derivatives
of $\phi(y)$ at $y=y_{jk}$ respectively.
We introduce Lagrange multiplier $\mu_{ij}$ for constraints (\ref{MPconstraint}). 
After optimizing the energy function of node $j$, 
the messages from node $j$ to $i$ are given by 
\begin{eqnarray}
\label{MPAB}
	&&A_{ij}\leftarrow-\mu_{ij},
	\\
\label{MPAB2}
	&&B_{ij}\leftarrow
	\cases{
	0 &for $h_{ij}^{-1}(0)>0$,\\  \\
	\displaystyle\biggl\{ \sum_{k\ne i} \cA_{jk}(R\phi''_{jk} + B_{jk})^{-1}
	\\
	\times\Theta\biggl[W-\biggr|y_{jk}
	-\frac{R\phi'_{jk}+A_{jk}+\mu_{ij}}{R\phi''_{jk}+B_{jk}}\biggr|\biggr]\biggr\}^{-1}
	&for $-\psi'(0)\leq h_{ij}^{-1}(0)\leq 0$,\\  \\
	\displaystyle\biggl\{ \psi''(\xi)^{-1} + \sum_{k\ne i} \cA_{jk}(R\phi''_{jk} + B_{jk})^{-1}
	\\
	\times\Theta\biggl[W-\biggr|y_{jk}
	-\frac{R\phi'_{jk}+A_{jk}+\mu_{ij}}{R\phi''_{jk}+B_{jk}}\biggr|\biggr]\biggr\}^{-1}
	&for $h_{ij}^{-1}(0)< -\psi'(0)$,\\  \\
	}
\end{eqnarray}
where 
\begin{eqnarray}
	g_{ij}(x)=[\psi'\circ h_{ij}](x) +x,
\end{eqnarray}
\begin{eqnarray}
\label{MPmu}
	\mu_{ij}=
	\cases{
	0  &for $h_{ij}^{-1}(0)>0$,\\
	h_{ij}^{-1}(0) &for $-\psi'(0)\leq h_{ij}^{-1}(0)\leq 0$,\\
	g_{ij}^{-1}(0) &for $ h_{ij}^{-1}(0)< -\psi'(0)$,\\
	}
\end{eqnarray}
and $h_{ij}(x)$ is defined by 
\begin{eqnarray}
\label{MPh}
	h_{ij}(x) = y_{ij} - \Lambda_j
	-\sum_{k \neq i}\max\biggl\{-W, \min\biggl[W, y_{jk} - \frac{R\phi'_{jk}
	+A_{jk}+x}{R\phi''_{jk}+B_{jk}}\biggr]\biggr\}
\end{eqnarray}
These variables can be interpreted as the cavity
variables of node $j$ in the absence of $i$,
when a current $y_{ij}$ is drawn from node $j$.
$\mu_{ij}$ is the cavity chemical potential of node $j$.
$h_{ij}(x)$ is the cavity shortage of resource at node $j$ when 
$\mu_{ij}$ takes the value $x$. 
$\psi'\circ h_{ij}(x)$ is then the corresponding dissatisfaction cost per unit resource
of node $j$.
The condition that $\psi'\circ h_{ij}(x)+x=0$ in \req{MPmu} is thus to require the chemical
potential to be set at the minus of the dissatisfaction cost per unit resource shortage.

For the frictionless case of $\psi'(0)=0$,
$\psi'(h)$ changes continuously when $h$ varies from zero to negative.
Hence when $y_{ij}$ changes,
$\psi'\circ h_{ij}$ changes continuously,
resulting in a continuous change in the chemical potential $\mu_{ij}$ as well as the 
first derivatives of the vertex free energy function $A_{ij}$.
Hence,
for the quadratic load balancing task,
defined by $\phi(y)=y^2/2$,
the vertex free energies in the recursion relation \req{recur} will be piecewise
quadratic with continuous slopes,
with respect to continuous changes of currents $y_{ij}$,
implying the consistency of the proposed optimization algorithm.
This validifies the replacement of the message functions by the two-parameter messages.

Since the messages are simplified to be the first two derivatives of the vertex
free energies,
it is essential for the nodes to determine the {\it working points} at which
the derivatives are taken.
That is,
each node needs to estimate the current $y_{ij}$ drawn by its ancestor.
This determination of the working point is achieved by passing additional 
information-provision messages among the nodes.
Here,
we describe the method of backward information-provision messages.
This is in contrast to conventional message-passing algorithms,
in which messages are passed in the forward direction only.
An alternative method of forward information-provision messages can also be formulated
following \cite{wong2006}.

After the minimization of the vertex free energy at a node $j$,
forward messages $(A_{ij}, B_{ij})$ are sent forward from node $j$ to ancestor
node $i$.
Optimal currents $y_{jk}$ are computed and sent backward from node $j$
to the descendent nodes $k\neq i$.
These backward messages serve as a key in information provision to descendents,
so that the derivatives in the subsequent messages are to be 
taken at the updated working points.
Minimizing the free energy (\ref{MPL}) with respect to $y_{jk}$, 
the backward message is found to be 
\begin{eqnarray}
\label{MPy}
	y_{jk}\leftarrow
	\max\biggl\{-W, \min\biggl[W, y_{jk} - \frac{R\phi'_{jk}
	+A_{jk}+\mu_{ij}}{R\phi''_{jk}+B_{jk}}\biggr]\biggr\}
\end{eqnarray}

When implemented on real networks,
the nodes are not divided into fixed generations,
and there are no fixed ancestors or descendents among the neighbors of a particular node.
Individual nodes are randomly chosen and updated,
by randomly setting one of its neighbors to be a temporary ancestor. 
This results in independent updates of the currents $y_{ij}$ and $y_{jk}$
in the opposite directions of the same link.
To achieve more efficient convergence of the algorithm,
we can set $y_{ij}=-y_{ji}$ during the updates.
For the cost function used in this work,
we find that this procedure speeds up the balance of demands from the two nodes connected
by the link.

An alternative distributed algorithm can be obtained by iterating
the chemical potentials of the nodes.
The optimal currents are given by \req{solution} in terms of the chemical potentials
$\mu_i$ which, 
from Eqs. (\ref{xi_define}) and (\ref{constraints}),
are related to their neighbors via 
\begin{eqnarray}
\label{CPmu}
	\mu_{i}=
	\cases{
	0 &for $h_{i}^{-1}(0)>0$,\\
	h_{i}^{-1}(0) &for $-\psi'(0)\leq h_{i}^{-1}(0)\leq 0$,\\
	g_{i}^{-1}(0) &for $ h_{i}^{-1}(0)< -\psi'(0)$,\\
	}
\end{eqnarray}
where $h_i(x)$ and $g_i(x)$ are given by 
\begin{eqnarray}
\label{CPhg}
	&&h_i(x) = -\Lambda_i-\sum_j\cA_{ij}Y(\mu_j-x),
	\nonumber\\
	&&g_i(x) = \psi'\circ h_i(x) + x,
\end{eqnarray}
with function $Y$ again given \req{solutionY}.
This provides a simple local iteration method for the optimization problem
in which the optimal currents can be evaluated from the potential
differences of neighboring nodes.
Simulation results in Section \ref{sec:sim} show that the 
chemical potential iterations have an excellent agreement with the 
message-passing algorithm in Eqs.(\ref{MPAB})-(\ref{MPh}).

We may interpret this algorithm as a price iteration scheme by noting that the 
Lagrangian in \req{Lagr} can be written as
\begin{eqnarray}
	L=\sum_{(ij)}\cA_{ij} L_{ij} + {\rm constant},
\end{eqnarray}
where
\begin{eqnarray}
	L_{ij} = \phi(y_{ij})+(\mu_i-\mu_j)y_{ij}.
\end{eqnarray}
Therefore the problem can be decomposed into independent optimization problems,
each for a current on a link.
$\mu_i$ is the storage price per unit resource at node $i$,
and each problem involves balancing the transportation cost on the link,
and the storage cost at node $i$ less that at node $j$.
From Eqs. (\ref{CPmu}) and (\ref{CPhg}),
we see that when node $i$ is short of resources,
$\mu_i=-\psi'\circ h(\mu_i)$.
That is,
the storage price at a node is equal to minus the dissatisfaction energy
of that node.
This means that the algorithm gives bonuses to the link controls 
instead of charging them,
so as to encourage them to station their resources at the nodes 
with resource shortage.
The amount of bonus per unit resource compensates exactly the 
dissatisfaction energy per unit resource.
This corresponds to a pricing scheme for the individual links to optimize,
which simultaneously optimizes the global performance. 
The concept is based on the same consideration as those used in distributed 
adaptive routing algorithms in communication networks \cite{kelly1991}.

\section{Results and Physical Implications}
\label{sec:sim}

We examine the properties of the optimized networks by
first solving the theoretical recursive equation (\ref{recur}) numerically
to obtain various quantities of interest,
including the energy, 
the fraction of idle links and saturated links.
The results are then compared with the simulation results obtained by the 
message-passing and price iteration algorithms.
All experiments assume a quadratic shortage cost
$\psi(\xi)=\xi^2/2$ for shortage $\xi$ given in Eq.~(\ref{xiCon}),
and a quadratic transportation cost $\phi(y)=y^2/2$,
and the capacity distribution $\rho(\Lambda)$ a Gaussian with mean 
$\avg{\Lambda}$ and variance 1.

To solve numerically the recursive equation (\ref{recur}), 
we have discretized the vertex free energy functions $F_V(y|{\mathbf T})$
into a vector, whose $i^{\rm th}$ component is the value of the
function corresponding to the current $y_i$.
At each generation of the calculation,
a node $j$ is randomly chosen and its capacity $\Lambda_j$ is drawn from 
the distribution $\rho(\Lambda_j)$.
The node is randomly connected to $c-1$ nodes of the previous generation
with vertex free energies  $F_V(y_k|{\mathbf T}_k)$.
Then the vertex free energy $F_V(y|{\mathbf T})$ is found by minimizing the 
right hand side of \req{recur} using standard numerical methods for 
every discretized components of $y$, 
subject to the constraints of $|y_k|\leq W$.
Thus a new generation of vertex energy functions is generated
and the process continue in a recursive manner.

\subsubsection{Effects of Bandwidths}

\begin{figure}
\centerline{\epsfig{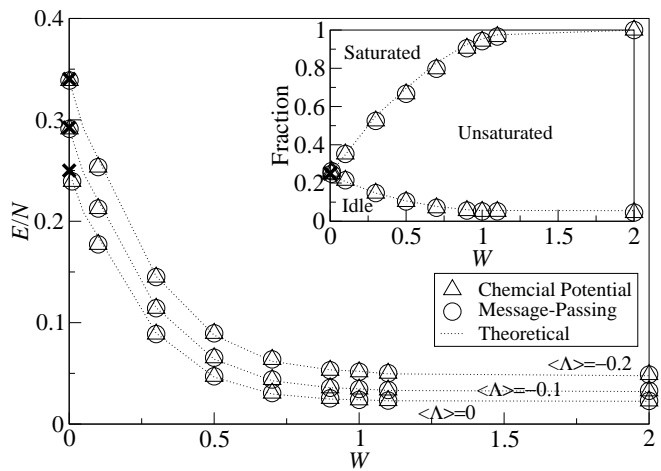}}
\caption{Simulation results of average energy per node as a function of bandwidth $W$
for 
$N=1000$, $c=3$, $R=0.1$ and $\langle\Lambda\rangle=0$, $-0.1$, $-0.2$
with 200 samples,
as compared to the numerical solution of Eq.~(\ref{recur}).
${\bf\times}$'s correspond to the energy in the zero bandwidth limit 
as obtained from Eq.~(\ref{E_W0}). 
Inset: The fraction of idle, unsaturated and saturated links as a 
function of bandwidth for $\langle\Lambda\rangle=$0. 
${\bf\times}$ corresponds to $f_{\rm idle}$ in the zero bandwidth limit 
as obtained from Eq.~(\ref{fidle_W0}). 
}
\label{varyW_smallC}
\end{figure}

Figure~\ref{varyW_smallC} shows the
average energy as a function of the bandwidth $W$,
obtained by,
respectively,
the message-passing and the price iteration algorithm simulations,
and the recursive equation (\ref{recur}).
Simulation results collapse almost perfectly with the recursive equation.
We find that
the average energy increases with decreasing bandwidths.
Physically, 
decreasing bandwidth corresponds to the increasing limitations in 
the allocation of resources.
Thus, it is natural to have higher energy.
In the limit of zero bandwidth with small finite connectivity, 
the solution is equivalent to the initially unoptimized condition.
As the bandwidth vanishes, 
no resources are transported among the network, 
and the average network energy is given by 
\begin{eqnarray}
\label{E_W0}
	E=N\int_{-\infty}^0 \frac{\Lambda^2}{2} \rho(\Lambda) d\Lambda.
\end{eqnarray}

In the limit of zero bandwidth,
a link can only be idle or saturated,
with no intermediate unsaturated state.
A link is idle only if the two connected nodes are satisfied already,
with their own initial resources.
The probability of finding an idle link is thus determined by the initial resources of 
the connected nodes,
yielding the fraction of idle links as 
\begin{eqnarray}
\label{fidle_W0}
	f_{\rm idle}=\left[\int_0^{\infty}\rho(\Lambda)d\Lambda\right]^2
	=\frac{1}{4}\left[1-{\rm erf}\left(\frac{\langle\Lambda\rangle}{\sqrt{2} }
	\right)\right]^2,
\end{eqnarray}
where the last equality is specific to the Gaussian $\rho(\Lambda)$.
The fraction of saturated link is indeed given by $1-f_{\rm idle}$.
The corresponding limit is shown in the inset of Fig.~\ref{varyW_smallC},
which is consistent with the simulation results.

\subsubsection{Bottleneck Effect}

Another interesting physical consequence of decreasing bandwidth is also
shown in the inset of Fig.~\ref{varyW_smallC}. 
We note that as the bandwidth decreases, 
the fraction of saturated link increases,
which is a direct consequence
of the attempt to minimize the shortages of nodes fed by the saturated links.
A similar agreement would lead us to anticipate a decreasing fraction of idle links as 
the bandwidth decreases, 
since more links should participate in the task of resource allocation.
Surprisingly, 
we notice an increasing fraction of idle links as the bandwidth decreases, 
in contrast with our anticipation.

\begin{figure}
\centerline{\epsfig{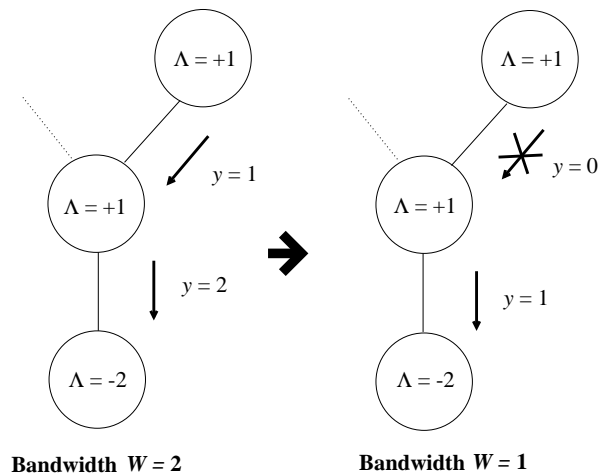}}
\caption{An example of bottleneck effect.}
\label{bottle}
\end{figure}

This is a consequence of the {\it bottleneck effect}, 
as illustrated in Fig.~\ref{bottle}. 
When the bandwidth decreases, 
resources transferred from the secondary neighbors 
may become redundant since resources from nearest neighbors already saturate 
the link to the dissatisfied node, 
which can therefore be considered as a bottleneck in transportation. 
This makes the resource distribution among the nodes less uniform, 
since resources are less free to be 
transported through the links. 
As a result, 
more nodes are having either excess resources or large shortage, 
leading to a higher network energy.
On the other hand,
the existence of these bottlenecks confirm the efficiency of our algorithms,
since redundant flows to distant nodes are eliminated.

The existence of bottlenecks is common in many real networks.
Among the most common examples are bottlenecks occurring in traffic congestions.
Further studies can be carried out to minimize the 
bottleneck effect,
such as using heterogeneous bandwidths for different links.
The highway is an example of enlarging the bandwidth on a link with
heavy traffic.
Similar considerations have been applied to routing of traffic on telecommunication
networks \cite{kelly1991}.

\subsubsection{Density of Saturated Links and Its Physical Implications}

\begin{figure}
\centerline{\epsfig{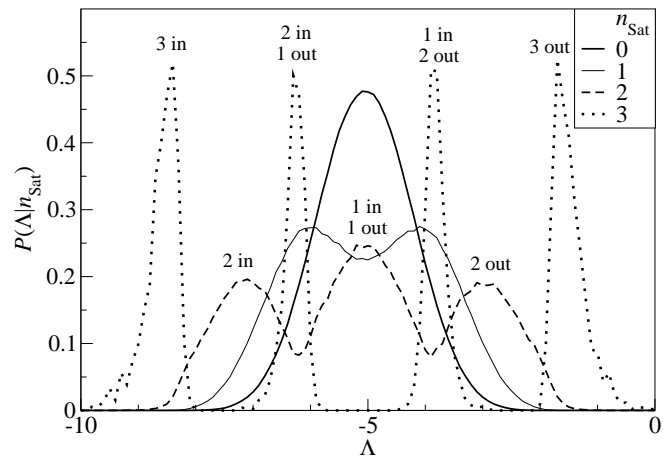}}
\caption{The distribution of initial resource $\Lambda$ given different number
of saturated links $n_{\rm Sat}$ after optimization for $N=1000$, 
$\langle\Lambda\rangle=-5$, $c=3$, $R=0.1$ 
and $W=1$ with 4000 samples.}
\label{satDis}
\end{figure}

To study the roles played by the different nodes in resource allocation,
we consider the example of networks with $c=3$ and very negative 
$\avg{\Lambda}$,
so that all resources in the networks have to be utilized for optimization.
We classify the nodes into four classes,
the nodes of each class having $n_{\rm sat}=0$ to 3 saturated links connected to them.
Figure \ref{satDis} shows the conditional capacity distribution of these four classes.

For nodes having no saturated links, 
their participation in the optimization task is relatively inactive,
and their capacity distribution is approximately Gaussian with 
an average $\langle\Lambda\rangle$.
For nodes having one saturated link, 
two peaks are found around $\Lambda-W$ and $\Lambda+W$ 
in the conditional capacity distribution,
respectively corresponding
to nodes acquiring resources from or providing resources to their neighbors, 
in order to lower the global shortage cost.

For nodes having two saturated links,
we obtain a trimodal distribution.  
The right and the left peaks at around $\Lambda+2W$ and $\Lambda-2W$
correspond to resource donors and receptors respectively,
which export or import resources, 
saturating two of the connected links.
The central peak at $\Lambda$ corresponds to nodes with initial resources 
close to the average value.
These nodes have little tendency to donate resources to their neighbors or receive from them.
Thus they are referred as {\it relays} which participate in optimization
by transferring resources from one neighbor to another.
For nodes with three saturated links, 
we observe a {\it four-modal} distribution.
Similarly, 
the rightmost and the leftmost peaks correspond to {\it sources}
and {\it sinks} of resources respectively.
The middle-right and 
the middle-left peaks correspond to {\it source-like relays} and 
{\it sink-like relays} respectively, 
namely,
they serve partially as relays and partially as sources and sinks.
All these distributions show the strong dependence of the number 
of saturated links of a node on its own initial resources, 
which identify an inborn optimization role for each node 
during the resource allocation.

\section{The High Connectivity Limit}
\label{sec:highC}

The optimal network behavior can be studied analytically in the limit of high connectivity.
In this limit,
the bandwidth of the links plays an important role in the scaling laws of the
physical quantities of interest.
Two cases will be considered in this section.
In the first case,
the bandwidths of the individual links remains constant when the connectivity scales up.
This increases the total available bandwidth connecting a node,
and hence the freedom in resource allocation.

In the second case, 
we scale the bandwidth of the individual links as $c^{-1}$ when the connectivity
$c$ scales up.
Hence,
the total available bandwidth connecting the individual nodes is conserved,
but the burden of transporting resources is divided into smaller currents
shared by a larger number of neighbors.

To further simplify the analysis, 
we set $\phi(y)=y^2/2$,
$\psi(\xi)=\xi^2/2$ and $R\sim O(1)$ in the 
subsequent derivations.

\subsection{High Connectivity with Non-vanishing Bandwidths}
\label{sec:nvb}

\subsubsection{The Optimal Solution}

In the high connectivity limit,
the magnitudes of the currents in individual links are reduced,
owing to the division of the transportation burden among the links.
Since the bandwidths remain constant,
the fraction of saturated links becomes negligible.
Hence,
Eq.~(\ref{MPAB2}) converges to the result
\begin{eqnarray}
\label{nvbB}
	B_{ij}=\Theta(-\mu_{ij})\frac{R}{c}.
\end{eqnarray}
At the steady state,
all current adjustments $\epsilon_{jk}$ vanish,
reducing Eq. (\ref{MPy}) to 
\begin{eqnarray}
\label{nvby}
	y_{jk}=\frac{1}{R}\left(A_{ij}-A_{jk}\right).
\end{eqnarray}
Thus,
Eq. (\ref{MPmu}) for $\mu_{ij}$ in the message-passing 
algorithms can be simplified to
\begin{eqnarray}
\label{nvbA}
	A_{ij}=
	\max\Biggl\{\Biggl(1+\frac{c}{R}\Biggr)^{-1}\biggl[y_{ij}-\Lambda_j+
	\frac{1}{R}\sum_{k\ne i}\cA_{jk}A_{jk}\biggr], 0\Biggr\},
\end{eqnarray}
where we have approximated $c-1$ by $c$.
This expression of $A_{ij}$ corresponds to the shortage on node $j$
after optimization for the quadratic cost $\psi'(\xi)=\xi$.
We then utilize the fact that the messages from the descendents are independent
to each other.
This allows us to express the collective effects
of the descendents on a node in terms of the statistical properties
of the descendents. 
By virtue of the law of large numbers,
it is sufficient to consider the mean $m_A$ and the variance $\sigma_A^2$
of the messages $A_{jk}$.
In Eq.~(\ref{nvbA}), 
the term $y_{ij}$ is negligible in the limit of high connectivity.
After applying the law of large numbers to the term $\sum_k\cA_{jk}A_{jk}$,
we can write
\begin{eqnarray}
\label{nvbA2}
	A_{ij}=\max\left(\frac{c m_A -R\Lambda_j}{c+R},0\right).
\end{eqnarray}
Averaging over $\Lambda_j$, 
which is drawn from a Gaussian distribution of mean
$\langle\Lambda\rangle$ and variance 1, 
we obtain a self-consistent expression for $m_A$, 
\begin{eqnarray}
\label{nvbSelf}
	&&\langle\Lambda\rangle+m_A=\frac{1}{\sqrt{2\pi}}
	\exp\left[
	-\frac{1}{2}\left(\frac{c m_A}{R}-\langle\Lambda\rangle\right)^2\right]
	\nonumber\\
	&&-\biggl(\frac{cm_A}{R}-\langle\Lambda\rangle\biggr)
	H\biggl(\frac{c m_A}{R}-\langle\Lambda\rangle\biggr),
\end{eqnarray}
where $H(x)=\frac{1}{2}[1-{\rm erf}(x/\sqrt{2})]$. 
In the limit of high connectivity,
this further implies
\begin{eqnarray}
\label{nvbmA}
	m_A=\max(-\langle\Lambda\rangle, 0).
\end{eqnarray}
This equation relates the average shortage $m_A$ after optimization to 
the average initial resources $\langle\Lambda\rangle$ of the system, 
implying a uniform resource redistribution in the 
large connectivity limit.
Given a negative $\langle\Lambda\rangle$, 
the equation implies no nodes are keeping excess resources
relative to the average $\langle\Lambda\rangle$.
All excess initial resources are redistributed to achieve a
global optimized state. 
We note that the currents scale as $(c+R)^{-1}$,
resulting in a much smaller transportation cost compared
with the shortage cost when the
connectivity increases.
This allows the re-allocation of initial resources through 
vanishing currents to minimize the shortage cost.
The sum of $c$ currents of $O[(c+R)^{-1}]$ results
in a total resource transport of $O(1)$ transported to or from a node.
It provides an extremely large freedom in resource allocation.
For positive $\langle\Lambda\rangle$,
the average shortage $m_A$ vanishes in the limit high connectivity limit,
corresponding to a zero fraction of unsatisfied nodes.
Hence,
increasing connectivity results in better resource allocation of the system.

\subsubsection{Network Properties}

We obtain the current distribution by combining the expression for $A_{ij}$ in
Eq.~(\ref{nvbA2}) with the equation for current in
Eq.~(\ref{nvby}),
such that 
\begin{eqnarray}
\label{nvby2}
        y_{ij} =\frac{1}{c+R}\biggl[\max\biggl(\frac{cm_A}{R}-\Lambda_i, 0\biggr)
        -\max\biggl(\frac{cm_A}{R}-\Lambda_j,0\biggr)\biggr].
\end{eqnarray}
Hence the current distribution is given by 
\begin{eqnarray}
\label{nvbpy}
        &&P(y)=\int d\Lambda_1\rho(\Lambda_1)
        \int d\Lambda _2\rho(\Lambda_2)\delta\Biggl\{y-
        \nonumber\\
        &&
        \frac{1}{c+R}\biggl|\max\biggl(\frac{cm_A}{R}-\Lambda_1, 0\biggr)
        -\max\biggl(\frac{cm_A}{R}-\Lambda_2, 0\biggr)
        \biggr|
        \Biggr\}.
\end{eqnarray}
For the Gaussian distribution $\rho(\Lambda)$, 
we obtain
\begin{eqnarray}
\label{nvbydis}
	&&P(y)=H^2(\chi)\delta(y)
	+\sqrt{\frac{2}{\pi}}(c+R)H(\chi)
	\exp\left[-\frac{1}{2}((c+R)y-\chi)^2\right]
	\\
	&&+\frac{c+R}{\sqrt{\pi}}H\biggl(\frac{(c+R)y-2\chi}{\sqrt{2}}\biggr)
	\exp\left[-\frac{1}{4}((c+R)^2y^2)\right],
	\nonumber
\end{eqnarray}
where $\chi\equiv c m_A/R-\langle\Lambda\rangle$.
This shows that the rescaled distribution $P[(c+R)y]/(c+R)$ is independent
of $c$ and solely dependent on $\langle\Lambda\rangle$.
This confirms that the currents scale as $(c+R)^{-1}$ as the
connectivity increases,
as described before.

We obtain the chemical potential distribution by 
considering the chemical potential of a central node 0 fed by $c$ descendents.
Introducing Lagrange multipliers,
the Lagrangian becomes
\begin{eqnarray}
\label{nvbL}
	&&L=\frac{\xi_{0}^2}{2}
	+\mu\biggl(\Lambda_0+\sum_{j=1}^c y_{0j}+\xi_{0}\biggr)
	+\alpha\xi_{0}
	+\sum_{j=1}^c\biggl[\frac{R y_{0j}^2}{2} + A_{0j}y_{0j}	
	\nonumber\\
	&&
	+\Theta(-\mu_{0j})\frac{R y_{0j}^2}{2c}
	+\lambda_{0j}^+(W-y_{0j})+\lambda_{0j}^-(W+y_{0j})\biggr].
\end{eqnarray}
The currents are given by $y_{0j}=(-A_{0j}-\mu)/R$,
and the chemical potential is given by 
\begin{eqnarray}
\label{nvbmu}
	\mu=\frac{1}{1+\frac{c}{R}}
	\min\biggl(\Lambda_0-\frac{1}{R}\sum_{j=1}^c A_{0j},0\biggr).
\end{eqnarray}
Using the statistical properties of $A_{0j}$ in the high connectivity limit,
the approximate expression for $\mu$ becomes
\begin{eqnarray}
\label{nvbmu2}
	\mu=\frac{1}{1+\frac{c}{R}}
	\min\biggl(\Lambda_0-\frac{c m_A}{R},0\biggr).
\end{eqnarray}
For Gaussian distribution of the initial resource,
we derive the chemical potential distribution given by 
\begin{eqnarray}
\label{nvbmudis}
	P(\mu)=\frac{1+\frac{c}{R}}{\sqrt{2\pi}}
	\exp\left\{-\frac{1}{2}
	\biggl[\biggl(1+\frac{c}{R}\biggr)\mu+\chi\biggr]^2\right\}\Theta(-\mu)
	+H\left(\chi\right)\delta(\mu).
\end{eqnarray}

Finally,
the average energy per node is given by
\begin{eqnarray}
\label{nvbavgE}
	\langle E\rangle=\frac{1}{2}\langle\mu^2\rangle
	+\frac{Rc}{4}\langle y^2\rangle,
\end{eqnarray}
where the first term corresponds to the shortage energy 
per node and the the second term corresponds to the 
transportation energy on the links connected to a node.
Using Eqs.~(\ref{nvby2}) and (\ref{nvbmu2}), 
the average energy per node becomes
\begin{eqnarray}
\label{nvbavgE1}
        \left\langle E \right\rangle
        =\frac{R}{2(c+R)}\biggl\langle
        \max\biggl(\frac{cm_A}{R}-\Lambda,0\biggr)^2\biggr\rangle
        -\frac{Rc}{2(c+R)^2}\biggl\langle
        \max\biggl(\frac{cm_A}{R}-\Lambda,0\biggr)
        \biggr\rangle^2.
        \nonumber\\
\end{eqnarray}
For the Gaussian initial resource distribution, this reduces to
\begin{equation}
\label{nvbavgE3}
        \left\langle E \right\rangle
        =\frac{R}{2(c+R)}I_2(\chi)
        -\frac{Rc}{2(c+R)^2}I_1(\chi)^2,
\end{equation}
where
\begin{equation}
\label{nvbI1}
        I_1 (\chi )=\int_{-\infty }^\chi {Dz(\chi -z)} =\frac{e^{-\frac{\chi
        ^2}{2}}}{\sqrt {2\pi } }+\chi H(-\chi ),
\end{equation}
\begin{equation}
\label{nvbI2}
        I_2 (\chi )=\int_{-\infty }^\chi {Dz(\chi -z)^2}
        =\chi \frac{e^{-\frac{\chi^2}{2}}}{\sqrt {2\pi } }
        +(\chi ^2+1)H(-\chi ).
\end{equation}

\subsubsection{Efficient Resource Allocation in The Shortage Reigme}

In the regime of negative $\langle\Lambda\rangle$,
apart from the uniform resource redistribution deduced from Eq.~(\ref{nvbmA}),
the current distribution,
the chemical potential distribution and the energy also show universal characteristics
of efficient resource allocation.
When $\langle\Lambda\rangle$ is negative,
$\chi\gg 1$ in the high connectivity limit,
the current distribution reduces to
\begin{eqnarray}
\label{nvbydis-}
	P(y)=\frac{c+R}{\sqrt{\pi}}
	\exp\left[-\frac{(c+R)^2y^2}{4}\right].
\end{eqnarray}
Physically,
the disappearance of the term $\delta(y)$ in 
the expression of the current distribution
for negative $\langle\Lambda\rangle$ implies that none of the nodes
are holding excess resources relative to the average
$\langle\Lambda\rangle$,
reproducing our conclusion from Eq.~(\ref{nvbmA}).

The chemical potential distribution becomes
\begin{eqnarray}
\label{nvbmudis-}
	P(\mu)=\frac{1+\frac{c}{R}}{\sqrt{2\pi}}
	\exp\left\{-\frac{1}{2}
	\biggl[\biggl(1+\frac{c}{R}\biggr)(\mu-\langle\Lambda\rangle)\biggr]^2\right\}.
\end{eqnarray}
This shows that the distribution $P[(1+c/R)(\mu-\langle\Lambda\rangle)]/(1+c/R)$
is independent of $c$ and depends solely on $\langle\Lambda\rangle$.
Thus, the width of the distribution scales as $(1+c/R)^{-1}$.
The variable $-\mu=\xi_0$,
corresponding to the shortage of a node.
This implies that the variance of the shortage distribution also scales as $(1+c/R)^{-1}$,
indicating a more 
efficient allocation of resources.
The absence of the delta function component at $\mu=0$ implies that none of the nodes
are holding excess resources in the negative $\langle\Lambda\rangle$ regime.

As both the current and chemical potential distributions
are Gaussian,
the average energy per node is simplified to
\begin{eqnarray}
\label{nvbavgE-}
	\langle E\rangle=\frac{1}{2}\langle\Lambda\rangle^2
	+\frac{R}{2(c+R)}.
\end{eqnarray}
Note that in the high connectivity limit, 
the average energy per node approaches $\langle\Lambda\rangle^2/2$.
This corresponds to the theoretical limit of efficient and uniform resource allocation,
in which the total shortage is evenly shared among all nodes.

\subsubsection{Comparison with Simulations}

\begin{figure}
\centerline{\epsfig{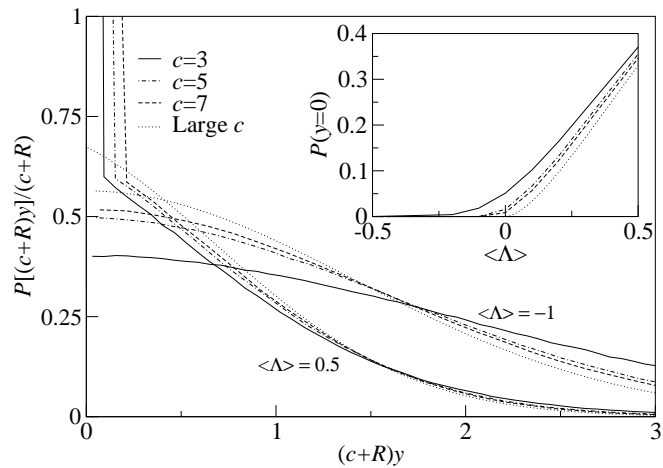}}
\caption{The rescaled current distribution $P[(c+R)y]/(c+R)$ 
for $N=10000$, 
$R=0.1$ and $W=1$ with 200 samples
at different values of $\langle\Lambda\rangle$, 
as compared to the corresponding theoretical large connectivity limits
given by Eqs.~(\ref{nvbydis}) and (\ref{nvbydis-}).
Inset:
The fraction of idle links $P(y=0)$ as a function of $\langle\Lambda\rangle$.
}
\label{gr_nvbydis}
\end{figure}

Simulation results are compared with the analytical prediction
in the high connectivity limit.
Figure~\ref{gr_nvbydis} shows the rescaled current distribution
$P[(c+R)y]/(c+R)$.
For positive $\langle\Lambda\rangle$,
the distributions for different connectivities collapse almost perfectly 
after the currents are rescaled by $(c+R)^{-1}$, 
with a very weak dependence on $c$,
approaching the high connectivity limit progressively.
For  negative $\avg{\Lambda}$,
the distributions also approach the high connectivity limit when the connectivity increases,
although the dependence on $c$ is larger.

The distribution has a delta function component at $y=0$
and a continuous component when $\langle\Lambda\rangle$ is positive.
However,
when $\langle\Lambda\rangle$ is negative, 
the delta function component disappears and the distribution 
approaches a Gaussian.
As shown in the inset of Fig.~\ref{gr_nvbydis},
the magnitude of the delta function component,
that is,
the fraction of idle links,
vanishes as $\langle\Lambda\rangle$
becomes more negative.
Another delta function component is observed at $|y|=W$
when $c$ is small,
and dissappears as $c$ increases when the distribution
approaches the high connectivity limit.

\begin{figure}
\centerline{\epsfig{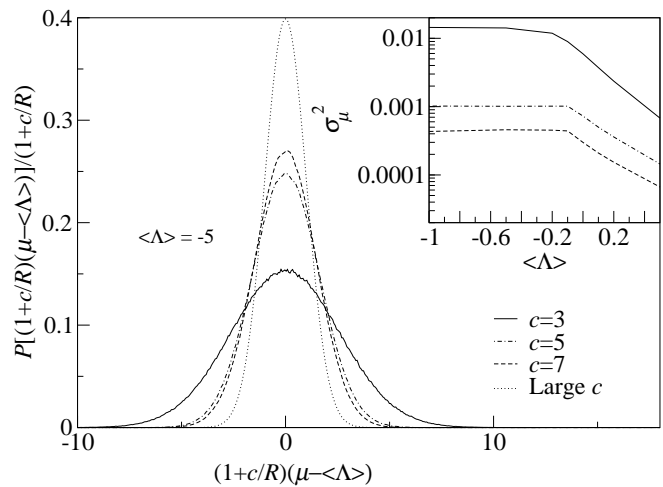}}
\caption{The rescaled chemical potential distribution $P[(1+c/R)(\mu-\langle\Lambda\rangle)]/(1+c/R)$ 
for $N=10000$, 
$R=0.1$, $W=1$ and $\langle\Lambda\rangle=-5$ with 200 samples, 
as compared to the theoretical large connectivity limit given by Eq.~(\ref{nvbmudis-}).
Inset:
The variance of chemical potential $\sigma_\mu^2$ as a function of $\langle\Lambda\rangle$.
}
\label{gr_nvbmudis}
\end{figure}

Figure~\ref{gr_nvbmudis} shows the rescaled chemical potential distributions
$P[(1+c/R)(\mu-\langle\Lambda\rangle)]/(1+c/R)$ compared with the high 
connectivity limit.
For negative $\langle\Lambda\rangle$, 
the rescaled distributions are Gaussian-like and approaching
the high connectivity limit. 
Compared with the corresponding
rescaled current distribution,
there is a larger dependence on $c$.
The inset of Fig. \ref{gr_nvbmudis} shows the variances of 
chemical potentials which approach different asymptotic values
values as $\langle\Lambda\rangle$ decreases,
for different fixed values of connectivity.
Remarkably,
for sufficiently negative $\langle\Lambda\rangle$,
the variances become $\langle\Lambda\rangle$
independent.
This further confirms the convergence of chemical potential distributions to 
universal Gaussian distributions in the negative
$\langle\Lambda\rangle$ regime.

\begin{figure}
\centerline{\epsfig{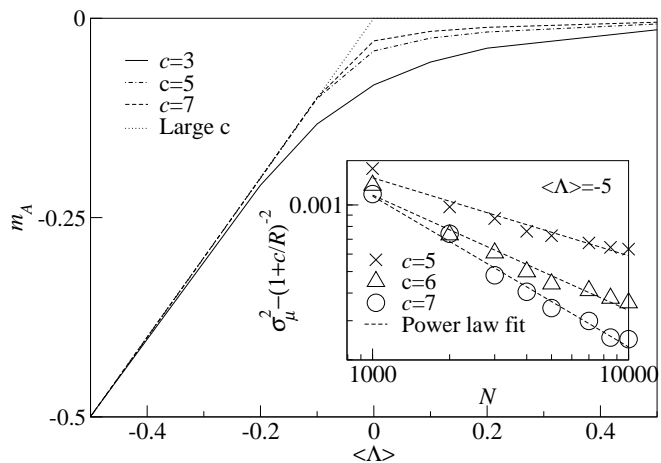}}
\caption{The parameter $m_A$ as a function of $\langle\Lambda\rangle$
for $N=10000$, 
$R=0.1$ and $W=1$ with 200 samples, 
as compared to the theoretical large connectivity limit
given by Eq.~(\ref{nvbmA}).
Inset:
The difference $\sigma_\mu^2-(1+c/R)^{-2}$
between the variance of the chemical potential and its corresponding large connectivity limit,
given by Eq. (\ref{nvbmudis-}) as a function of system size $N$. 
The dashed lines shows the power law fits.
}
\label{gr_nvbma}
\end{figure}

Figure~\ref{gr_nvbma} shows the average shortage,
as measured by $m_A$,
obtained from simulations at different connectivities as compared to the theory.
In the neighborhood of $\langle\Lambda\rangle\approx 0$,
as the connectivity increases,
$m_A$ gradually approaches the high
connectivity limit which corresponds to most uniform allocation 
of resources given by \req{nvbmA}.
Away from $\langle\Lambda\rangle\approx 0$,
$m_A$ collapses well with the analytical results.

Remarkably, 
the distributions of currents and chemical potentials, 
and the average shortage $m_A$,
all approach their high connectivity limits, 
for relatively low values of $c$ already.
We also analyze the finite size effects of the system in approaching 
the asymptotic behavior of the high connectivity limit. 
The inset of Fig.~\ref{gr_nvbma} shows the difference between the variance of 
chemical potentials and its asymptotic value $(1+c/R)^{-2}$
When the connectivity is low,
the differences are larger as indicated by 
the power law fit.

To summarize,
our analysis shows that in the high connectivity limit,
the optimized state of the system is one in which the resources are 
uniformly allocated,
such that the final resources on every individual node are equal.
Simulations with increasing connectivities reveal
the asymptotic approach to this uniform limit,
and the deviation from the limit decreases as the connectivity and the system size increase.

\subsection{High Connectivity with Vanishing Bandwidth}

We now consider the case that the bandwidth of individual links scales as 
$\tilde W/c$ when the connectivity increases,
where $\tilde W$ is a constant. 
Thus the total bandwidth $\tilde W$ available to an individual node remains 
a constant.
This is applicable to real networks in which the allocation of resources
is limited by the processing power of individual nodes.

\subsubsection{The Well-defined Chemical Potential Function}

We start by writing the chemical potentials using \req{CPmu},
\begin{eqnarray}
\label{vbmu}
	\mu_i=\min\biggl[\Lambda_i
	+\sum_{j=1}^N\cA_{ij}Y(\mu_j-\mu_i),0\biggr].
\end{eqnarray}
In the high connectivity limit,
the interaction of a node with all its connected neighbors become self-averaging,
making it a function which is singly dependent on its own chemical potential,
namely,
\begin{eqnarray}
\label{vbMmu}
	\sum_{j=1}^N\cA_{ij}Y(\mu_j-\mu_i)\approx c M(\mu_i).
\end{eqnarray}
Physically,
the function $M(\mu)$ corresponds to the average interaction of a node
with its neighbors when its chemical potential is $\mu$.
Thus,
we can write Eq.~(\ref{vbmu}) as
\begin{eqnarray}
\label{vbmu1a}
	\mu=\min[\Lambda+c M(\mu),0],
\end{eqnarray}
where $\mu$ is now a function of $\Lambda$,
and we have 
\begin{eqnarray}
\label{vbMmuint1}
	M(\mu_i)
	=\int_{-\infty}^\infty d\Lambda\rho(\Lambda)
	Y(\mu(\Lambda)-\mu_i)
\end{eqnarray}
where we have written the chemical potential of the neighbors as $\mu(\Lambda)$,
assuming that they are well-defined functions of their capacities $\Lambda$.

To explicitly derive $M(\mu)$, 
we take advantage of the fact that the rescaled bandwidth, 
$\tilde W/c$ vanishes in the high connectivity limit,
so that the current function $Y(\mu_j-\mu_i)$ is effectively a sign function,
which implies that the current on a link is always saturated.
(This approximation is not fully valid if $c$ is large but finite,
and will be further refined in subsequent discussions.)
Thus, 
we approximate
\begin{eqnarray}
\label{vbMmuint2}
	M(\mu_i)
	=\frac{\tilde W}{c}\int_{-\infty}^\infty d\Lambda\rho(\Lambda)
	{\rm sgn}[\mu(\Lambda)-\mu_i].
\end{eqnarray}
Assuming that $\mu(\Lambda)$ is a monotonic function of $\Lambda$,
then we have ${\rm sgn}[\mu(\Lambda)-\mu_i]={\rm sgn}(\Lambda-\Lambda_i)$,
such that
\begin{eqnarray}
\label{vbMmuint3}
	M(\mu_i)
	=\frac{\tilde W}{c}\Biggl[\int_{\Lambda_i}^\infty d\Lambda\rho(\Lambda)
	-\int_{-\infty}^{\Lambda_i} d\Lambda\rho(\Lambda)\Biggr].
\end{eqnarray}
For Gaussian distribution of capacities,
$M(\mu_i)$ is given by 
\begin{eqnarray}
\label{vbMmuint4}
	M(\mu_i)
	=-\frac{\tilde W}{c}{\rm erf}\biggl(
	\frac{\Lambda_i-\langle\Lambda\rangle}{\sqrt{2}}\biggr).
\end{eqnarray}
Thus, $\mu(\Lambda)$ is explicitly given by 
\begin{eqnarray}
\label{vbmu2}
	\mu=\min\biggl[\Lambda-\tilde W{\rm erf}\biggl(
	\frac{\Lambda-\langle\Lambda\rangle}{\sqrt{2}}\biggr),0\biggr].
\end{eqnarray}
This equation relates the chemical potential of a node, 
i.e. the shortage after resource allocation, 
to its initial resource before.
Resource allocation through a large number of links with 
a fixed total bandwidth results in a well-defined function relating the two quantities.

\subsubsection{Maxwell's Construction}

\begin{figure}
\centerline{\epsfig{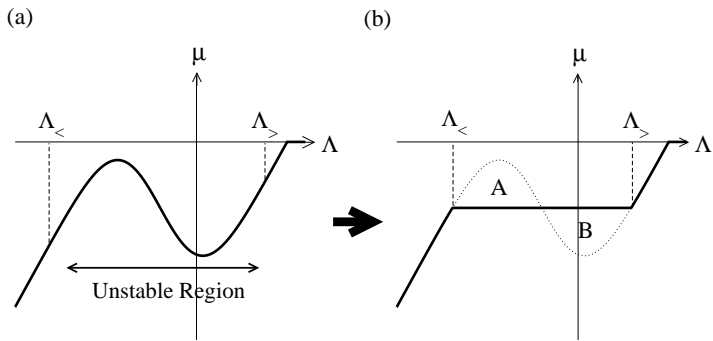}}
\caption{Maxwell's construction on $\mu(\Lambda)$.}
\label{gr_maxwell}
\end{figure}

\req{vbmu2} gives a well-defined function $\mu(\Lambda)$ as long as
$\tilde W\leq \sqrt{\pi/2}$.
However,
a phase transition takes place when $\tilde W$ reaches $\sqrt{\pi/2}$.
When $\tilde W> \sqrt{\pi/2}$,
turning points exists in $\mu(\Lambda)$ as shown in Fig.~\ref{gr_maxwell}(a), 
due to the dominant effect of the error function component in the equation.
$\mu(\Lambda)$ is no longer a monotonic function.
This creates a thermodynamically unstable scenario,
since in the region of $\mu(\Lambda)$ with negative slope,
nodes with lower capacities have higher chemical potentials than their neighbors with 
higher capacities.
This implies that the current flows from poorer nodes to richer ones.
Mathematically,
the non-monotonicity of $\mu(\Lambda)$ means that ${\rm sgn}[\mu(\Lambda)-\mu_i]$ 
and ${\rm sgn}(\Lambda-\Lambda_i)$ are no longer necessarily equal,
and \req{vbmu2} is no longer valid.

Nevertheless,
\req{vbmu} permits another solution of constant $\mu$ in a range of $\Lambda$.
This is possible since for any $\mu_i$ and $\mu_j$ in this range,
$|\mu_i-\mu_j|\leq R\tilde W/c$ implies that the link between nodes $i$ and $j$ is unsaturated.
If an extensive fraction of the $c$ links of such a node are connected to other nodes in the range,
then the freedom of tuning the currents in the unsaturated
links enables the nodes in the range to have the same level of resources after
optimization.

Hence,
we propose that the unstable region of $\mu(\Lambda)$ should be 
replaced by a range of constant $\mu$ as shown in Fig.~\ref{gr_maxwell}(b)
analogous to Maxwell's construction in thermodynamics.
Let $(\Lambda_<, \mu_o)$ and $(\Lambda_>, \mu_o)$ be the end points of 
the Maxwell's construction as shown in Fig.~\ref{gr_maxwell}(b).
The position of this construction can be determined by the conservation
of resources.
Since the construction is not necessary for positive $\avg{\Lambda}$,
we focus on the case of negative $\avg{\Lambda}$.
In the high connectivity limit,
resources are so efficiently allocated that the resources of the rich nodes
are maximally allocated to the poor nodes as much as the total
bandwidth $\tilde W$ allows.
Let $\Lambda_o$ be the capacity of a node beyond which nodes can 
have positive final resources 
(i.e. zero chemical potential).
Nodes with $\Lambda\geq\Lambda_o$ send out their resources without drawing
inward currents from their neighbors,
and can be regarded as {\it donors}.
The value of $\Lambda_o$ is given by the self-consistent equation
\begin{eqnarray}
\label{vblambdao}
	\Lambda_o=\tilde W\int_{-\infty}^{\Lambda_o} d\Lambda\rho(\Lambda).
\end{eqnarray}
For Gaussian distribution of capacities,
\begin{eqnarray}
	\Lambda_o=\frac{\tilde W}{2}\biggl[1+{\rm erf}\biggl(
	\frac{\Lambda_o-\langle\Lambda\rangle}{\sqrt{2}}\biggr)\biggr],
\end{eqnarray}
By equating the networkwide shortage to minus the initially 
available resources,
we find 
\begin{eqnarray}
\label{vbconstraint}
	-\sum_i\mu_i=\sum_i\biggl[-\Lambda_i\Theta(\Lambda_o-\Lambda_i)
	-\Lambda_o\Theta(\Lambda_i-\Lambda_o)\biggr].
\end{eqnarray}
On the right hand side,
the first term corresponds to the sharing of shortages among
the non-donors,
while the second term corresponds to the resources donated by the donors
to reduce the total shortage.
In the high connectivity limit,
this equation becomes
\begin{eqnarray}
	-\int_{\Lambda_<}^{\Lambda_>} d\Lambda\rho(\Lambda)\mu_o
	-\Biggl(\int_{\Lambda_>}^{\Lambda_o} +\int_{-\infty}^{\Lambda_<}\Biggr)
	d\Lambda\rho(\Lambda)\mu(\Lambda)
	\nonumber\\
	=-\int_{-\infty}^{\Lambda_o} d\Lambda\rho(\Lambda)\Lambda
	-\Lambda_o\int_{\Lambda_o}^{\infty} d\Lambda\rho(\Lambda).
\end{eqnarray}
Substituting Eqs. (\ref{vbmu1a}), (\ref{vbMmuint3}) and (\ref{vblambdao}), 
and making use of the symmetry relation
\begin{eqnarray}
	\tilde W\int_{\infty}^{\Lambda_o}d\Lambda\rho(\Lambda)
	\int_{\infty}^{\Lambda_o}d\Lambda'\rho(\Lambda'){\rm sgn}(\Lambda'-\Lambda) = 0
\end{eqnarray}
we arrive at the condition
\begin{eqnarray}
	\mu_o\int_{\Lambda_<}^{\Lambda_>} d\Lambda\rho(\Lambda)
	= \int_{\Lambda_<}^{\Lambda_>} d\Lambda\rho(\Lambda)\mu(\Lambda),
\end{eqnarray}
which implies that the value of $\mu_o$ should be chosen such that
the areas A and B in Fig.~\ref{gr_maxwell}(b),
weighted by the distribution $\rho(\Lambda)$,
should be equal -- in 
direct correspondence with the Maxwell's construction in thermodynamics.

For capacity distributions $\rho(\Lambda)$ symmetric with respect to $\avg{\Lambda}$,
we have
\begin{eqnarray}
	\mu_o = \avg{\Lambda} = \frac{1}{2}(\Lambda_<+\Lambda_>).
\end{eqnarray}
As a result, 
the function $\mu(\Lambda)$ is given by 
\begin{eqnarray}
\label{vbmuhori}
	\mu(\Lambda) = 
	\cases{
  	\langle\Lambda\rangle   &for $\mu_<<\mu<\mu_>$,\\
  	\min\left[\Lambda-
  	\tilde W{\rm erf}\left(\frac{\Lambda-\langle\Lambda\rangle}{\sqrt{2}}\right),0\right]
  	&for otherwise,\\
	}
\end{eqnarray}
where as $\Lambda_<$ and $\Lambda_>$ are respectively given by the lesser and 
greater roots of the equation
\begin{eqnarray}
\label{vblambdaless}
	x=\langle\Lambda\rangle
	+\tilde W{\rm erf}\biggr(\frac{x-\langle\Lambda\rangle}{\sqrt{2}}\biggl).
\end{eqnarray}

Nodes $i$ with chemical potentials $\mu_i=\avg{\Lambda}$ represent clusters
of nodes interconnected by an extensive fraction of unsaturated links,
which provides the freedom to fine tune their currents so that
the shortages among the nodes are uniform.
They will be referred to as the {\it balanced} nodes.
On the other hand,
nodes outside the balanced clusters are connected by saturated links only.

\begin{figure}
\centerline{\epsfig{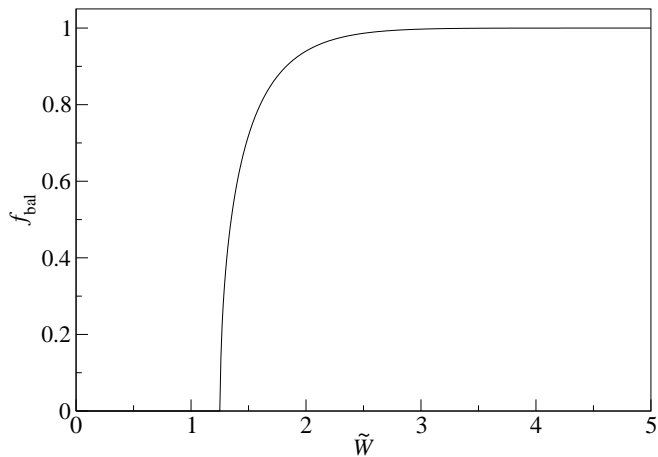}}
\caption{The dependence of the fraction of balanced nodes on the
bandwidth $\tilde W$.}
\label{fbal}
\end{figure}

The fraction $f_{\rm bal}$ of balanced nodes is given by the equation
\begin{eqnarray}
	f_{\rm bal}={\rm erf}\biggl(\frac{\tilde W f_{\rm bal}}{\sqrt{2}}\biggl).
\end{eqnarray}
Note that $f_{\rm bal}$ has the same dependence on $\tilde W$ for all negative $\avg{\Lambda}$.
Fig.~\ref{fbal} shows that when
the total bandwidth $\tilde W$ increases,
the fraction of balanced nodes increases,
reflecting the more efficient resource allocation brought by the 
convenience of increased bandwidths.
When $\tilde W$ becomes very large,
a uniform chemical potential of $\avg{\Lambda}$ networkwide is recovered,
converging to the case of non-vanishing bandwidths.

The chemical potential distribution can be obtained from Eq.~(\ref{vbmuhori}). 
Unlike the corresponding distribution in the case of non-vanishing bandwidths
described in the previous subsection, 
there is no explicit scaling on the connectivity.
In addition,
it is not purely Gaussian.
The delta function component $\delta(\mu-\avg{\Lambda})$
corresponds to the balanced clusters.
The weight of the delta function component is a measure of the efficiency
of resource allocation.

\begin{figure}
\centerline{\epsfig{figure=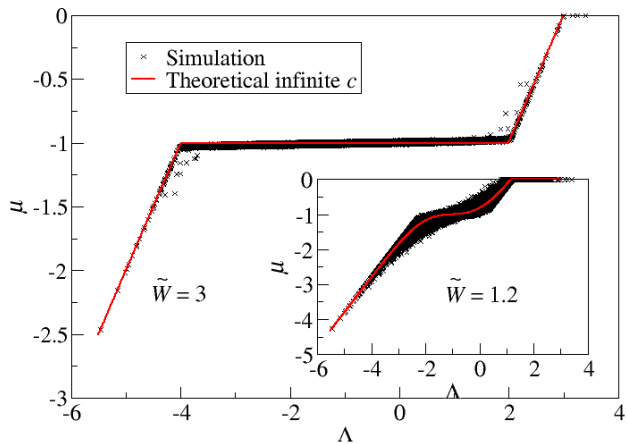, width=0.6\linewidth}}
\caption{(Colour online) The simulation results of $\mu(\Lambda)$ 
for $N=10000$, $c=15$, $R=0.1$, $\langle\Lambda\rangle = -1$ and $\tilde W=3$
with 70000 data points, 
compared with theoretical prediction.
Inset: The corresponding results for $\tilde W=1.2$.}
\label{gr_vbhori}
\end{figure}

We compare the analytical result of $\mu(\Lambda)$ in Eq.~(\ref{vbmuhori})
with simulations 
in Fig.~\ref{gr_vbhori}.
For $\tilde W > \sqrt{\pi/2}$, 
data points $(\Lambda, \mu)$ of individual nodes from network simulations 
follow the analytical result
of $\mu(\Lambda)$, 
giving an almost perfect overlap of data.
The presence of the balanced nodes with effectively constant chemical potentials 
is obvious and essential to explain the behavior of the majority of data
points from simulations.
Outside the region of balanced nodes,
the data points follow the tails of the function $\mu(\Lambda)$.
For $\tilde W<\sqrt{\pi/2}$, 
the analytical $\mu(\Lambda)$ shows no turning point as shown in 
the inset of Fig.~\ref{gr_vbhori}.
Despite the scattering of data points,
they generally follow the trend 
of the theoretical $\mu(\Lambda)$. 
This scattering effect may be explained by the use of a finite connectivity
in simulations. 
We found that the extent of scattering can be much reduced by increasing
the connectivity in the simulations.

\subsubsection{Large but Finite Connectivity}

Our analysis can be generalized to the case of large but finite connectivity, 
where the approximation in \req{vbMmuint2} is not fully valid.
This modifies the chemical potentials of the balanced nodes,
for which \req{vbMmuint2} has to be replaced by 
\begin{eqnarray}
\label{vbslantMmu}
	M(\mu)=
	\frac{\tilde W}{c}\biggl[
	\int_{\Lambda_>}^{\infty}d\Lambda\rho(\Lambda)
	-\int_{-\infty}^{\Lambda_<}d\Lambda\rho(\Lambda)\biggr]
	+\int_{\Lambda_<}^{\Lambda_>}d\Lambda\rho(\Lambda)
	\biggl(\frac{\mu(\Lambda)-\mu}{R}\biggr).
	\nonumber\\
\end{eqnarray}
We introduce an ansatz of a linear relationship between
$\mu$ and $\Lambda$ for the balanced nodes,
namely,
\begin{eqnarray}
\label{vbslantanastz}
	\mu=m\Lambda+b.
\end{eqnarray}
After direct substitution of \req{vbslantanastz} into $M(\mu)$ given by 
\req{vbslantMmu},
we get the self-consistent equations for $m$ and $b$,
\begin{eqnarray}
\label{vbslantm}
	m=\frac{R}{R+c~{\rm erf}\left(
	\frac{\Lambda_>-\langle\Lambda\rangle}{\sqrt{2}}\right)},
	\\
\label{vbslantb}
	b=\frac{c~{\rm erf}\left(
	\frac{\Lambda_>-\langle\Lambda\rangle}{\sqrt{2}}\right)}
	{R+c~{\rm erf}\left(
	\frac{\Lambda_>-\langle\Lambda\rangle}{\sqrt{2}}\right)}\langle\Lambda\rangle.
\end{eqnarray}
Thus,
the Maxwell's construction has a non-zero slope when the connectivity is finite.

We remark that the approximation in \req{vbslantMmu} assumes that the 
potential differences of the balanced nodes lie in the range of $2R\tilde W/c$,
so that their connecting links remain unsaturated.
Note that the end points of the Maxwell's construction have chemical potentials
$\avg{\Lambda}\pm R\tilde W/c$ respectively,
rendering the approximation in \req{vbslantMmu} {\it exact} at one special point,
namely,
the central point of the Maxwell's construction. 
Hence,
this approximation works well in the central region of the Maxwell's construction.
However,
when one approaches the end points of the Maxwell's construction,
the balanced nodes are also connected to nodes outside the balanced clusters with 
potential differences less than $2 R\tilde W/c$.
Hence,
we expect to see deviations from the theoretical 
linear prediction near the end points of the Maxwell's construction.

The chemical potential distribution can be obtained.
Compared with corresponding distribution in the high connectivity limit,
the delta function component of the balanced nodes is now smeared into 
a Gaussian component lying in a strip of width $m(\Lambda_>-\Lambda_<)$.
It implies a lower efficiency in resource allocation due 
to the finiteness of the connectivity.

\begin{figure}
\centerline{\epsfig{figure=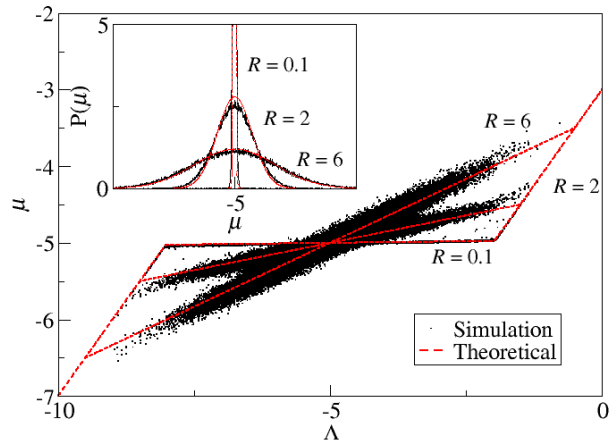, width=0.6\linewidth}}
\caption{(Colour online) Simulation results of $(\Lambda, \mu)$
for $N=2\times10^5$, $\tilde W=3$, $c=12$ and $\langle\Lambda\rangle=-5$
at different values of $R$, each with 65000 data points.
as compared to the theoretical predictions. 
Inset: the corresponding chemical potential distribution $P(\mu)$ of the 3 cases.
}
\label{gr_vbslant}
\end{figure}

In the simulation data shown in
Fig.~\ref{gr_vbslant},
the data points of $(\Lambda, \mu)$ from different resistance $R$
follow the trend of the corresponding analytical results, 
both within and outside the linear region,
with increasing scattering within the linear region as $R$ increases. 
As expected, 
there are derivations between the analytical and simulational results
at the two ends of the linear region,
with smoothened corners appearing in the simulation data,
especially in the case of $R=2$.
We note that when $R$ increases,
the gradient of the linear region increases, 
corresponding to a less uniform allocation of resources caused by higher
transportation costs during optimization.
As shown in the inset of Fig.~\ref{gr_vbslant}, 
the chemical potential distributions follow the trend of the analytical results.

Remarkably, 
as evident from Eq.~(\ref{vbslantm}),
even with constant available bandwidth $\tilde W$, 
increasing connectivity causes $m$ to decrease,
and hence sharpens the chemical potential distribution.
The narrower distributions correspond to higher efficiency in resource allocation.
It leads us to realize the potential benefits of 
increasing connectivity in network
optimization even for a given constant total bandwidth connecting a node, 
despite a decrease in bandwidth on individual links.

\subsubsection{Resource Allocation in Scale-free Networks}

Recent studies show that complex communication networks have highly 
hetergeneous structures,
and the connectivity distribution obeys a power law \cite{barabasi}.
These networks are commonly known as scale-free networks
and are characterized by the presence of hubs.
The presence of these nodes
with very high connectivity 
can modify the network behavior significantly.

\begin{figure}
\centerline{\epsfig{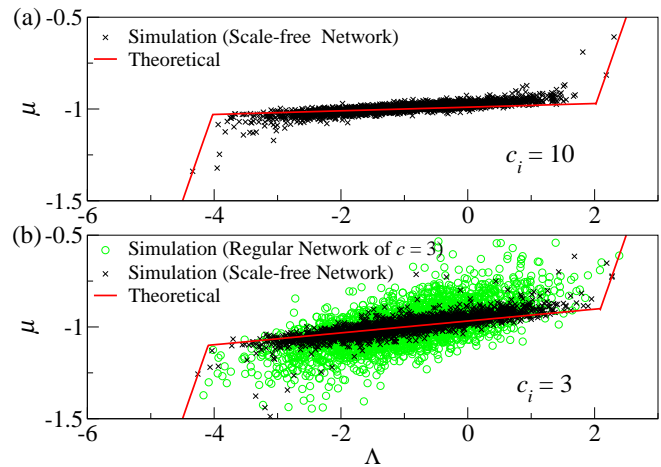}}
\caption{(Colour online) Simulation results of $(\Lambda, \mu)$
for $N=2\times10^5$, $R=0.1$, $\tilde W=3$ and $\langle\Lambda\rangle=-1$
as compared with theoretical results,
for 
(a) nodes with $c_i=10$ in scale-free networks with $P(c_i)\sim c_i^{-3}$, 
(b) nodes with $c_i=3$ in scale-free networks
and nodes in regular networks with $c=3$.
Each data set contains 2500 data points.
}
\label{scaleFree1}
\end{figure}

Figure \ref{scaleFree1} shows the simulation results of node $i$ with $c_i=3$
and $c_i=10$ in a scale-free network,
where $c_i$ is the connectivity of node $i$ and is drawn from the distribution $P(c_i)\sim c_i^{-3}$
when the scale-free network is constructed.
The data points of $(\Lambda, \mu)$ follow the 
corresponding analytical results of Eqs. (\ref{vbslantm}) and (\ref{vbslantb}),
for both sets of nodes with $c_i=3$ and $c_i=10$.
This implies that the previous argument of increasing efficiency by 
increasing connectivity also hold 
for scale-free networks,
as a smaller gradient $m$ is found for nodes with higher connectivity.
As can be seen from Fig. \ref{scaleFree1},
the data points are less scattered for nodes with large $c_i$,
implying a more efficient resource allocation for the hubs,
in addition to the effect of smaller $m$.

More important,
nodes with low connectivity benefit from
the presence of hubs in the networks.
To see these benefits,
the simulation results of nodes in scale-free networks are compared with
nodes in regular networks of the same connectivity.
As shown in Fig. \ref{scaleFree1}(b),
the data points from regular networks
are more scattered away from the Maxwell's construction,
when compared with those from scale-free networks.
This shows that
the presence of hubs increases the efficiency of the entire network,
especially for nodes with low connectivity.
This provides support that
scale-free networks are better candidates for resource allocation
than regular networks.

\section{Conclusion}
\label{sec:Conclusion}

We have applied statistical mechanics to a system in which magnitude of the interactions
(such as currents in resource allocation)
between on-site variables 
(such as the shortages) are constrained.
This allows us to study an optimization task of resource allocation on a sparse network,
in which nodes with different capacities are connected 
by links of finite bandwidths.
Analogy can be drawn between interaction of spins 
in magnetic systems and that of resources of nodes in networks. 
The bandwidths serve as constraints on the interaction magnitudes
and limit the information exchange among the neighbors.
Despite these additional constraints,
we found that the analyses of the unrestricted case in \cite{wong2006, wong2007} 
are applicable after appropriate adaptations,
such as allowing for shortages with finite penalty and obtaining
a non-linear relationship between currents and potential differences.

By adopting suitable cost functions, 
such as quadratic transportation and shortage costs,
the model can be applied to the study of realistic networks
with constrained transport between neighbors.
In this paper,
we focus on cases in which the shortage dominates.
By employing the Bethe approximation or equivalently the replica method, 
recursive relations of the vertex free energies can be derived.

Analytically,
the recursive relations enable us to make theoretical predictions
of various quantities of interest,
including the average energy,
the current distribution,
and the chemical potential distribution.
The predictions are confirmed by simulation results.
In particular,
the study reveals interesting effects due to finite
bandwidths.
When the bandwidth decreases,
resource allocation is less efficient,
and links are more prone to saturation.
A consequence is the creation of bottlenecks,
which refer to the saturation of links feeding the poor nodes,
rendering the secondary provision of resources from their next nearest neighbors redundant.
This causes certain links previously assigned for secondary transport to become idle,
lowering the participation of individual links in global optimization and making
the resources less uniformly distributed.
In the context of resource allocation,
further studies can be carried out to suppress the bottleneck effect.

An equally remarkable phenomenon is found in networks with fixed total bandwidths per node,
where bandwidths per link vanish in the high connectivity limit
and the relation between the chemical potential and capacity is well defined.
For sufficiently large total bandwidths,
we find a phase transition beyond which
the chemical potential function has to be described by the Maxwell's
construction,
implying the existence of clusters of balanced nodes having a uniform shortage among them.
In the case of large but finite connectivity,
the Maxwell's construction becomes a linear region with nonzero slope,
implying a less uniform shortage among the balanced nodes.
This reflects the benefits of increasing the number of connections in resource allocation.
When adapted to scaled-free networks,
deviations from the Maxwell's construction reveal that the presence
of hubs is able to homogenize the resources among 
the nodes with low connectivities.

For future extensions,
the theory and algorithms in this paper can be generalized to model real networks
with inhomogeneous connectivity and bandwidths.
Scale-free networks with adjustable bandwidth distributions would be
one of the most interesting systems to study.
Further studies can also be done on minimizing the bottleneck effect by using heterogeneous 
bandwidths on different links.
It is also worthwhile to consider other non-linear shortage costs.
We believe that the techniques presented in this paper are useful in 
many different network optimization problems and will lead to a large variety of potential applications.

\section*{Acknowledgements}
We thank David Saad for very meaningful discussions.
This work is supported
by the Research Grant Council of Hong Kong
(grant numbers HKUST 603606 and HKUST 603607).



\section*{References}
\begin{thebibliography}{10}

\bibitem{hertz}
J. Hertz, A. Krogh, and R. G. Palmer, 1999
{\it Introduction to the Theory of Neural Computation}
(Redwood City: Addison-Wesley)

\bibitem{nishimori} 
H.~Nishimori, 2001
{\it Statistical Physics of Spin Glasses and Information Processing}
(Oxford, UK: Oxford University Press)

\bibitem{challet}
D. Challet, M. Marsili and Y.-C. Zhang, 2005
{\it Minority Games}
(Oxford, UK: Oxford University Press)

\bibitem{kabashima} 
Y.~Kabashima and D.~Saad, 
{\it Statistical mechanics of low-density parity-check codes},
2004 J.~Phys.~A {\bf 37} R1

\bibitem{wong2006} 
K.~Y~.M.~Wong and D.~Saad, 
{\it Equilibration through local information exchange in networks},
2006 Phys.~Rev.~E {\bf 74} 010104(R)

\bibitem{wong2007}
K.~Y~.M.~Wong and D.~Saad, 
{\it Inference and optimization of real edges on sparse graphs: A statistical physics perspective},
2007 Phys.~Rev.~E {\bf 76} 011115

\bibitem{peterson} 
L.~Peterson and B.S.~Davie, 
2000 {\it Computer Networks: A Systems Approach} 
(San Diego CA: Academic Press)

\bibitem{ho}
Y. C. Ho, L. Servi, and R. Suri,
{\it A class of center-free resource allocation algorithms},
1980 Large Scale Syst. {\bf 1} 51

\bibitem{shenker}
S. Shenker, D. Clark, D. Estrin and S. Herzog,
{\it Pricing in computer networks: Reshaping the research agenda},
1996 Comput. Commun. Rev. {\bf 26} 19 

\bibitem{rardin}
R. L. Rardin,
1998 {\it Optimization in Operations Research}
(Englewood Cliffs, NJ: Prentice Hall)

\bibitem{bohn} 
S.~Bohn and Marcelo O.~Magnasco, 
{\it Structure, Scaling, and Phase Transition in the Optimal Transport Network},
2007 Phys. Rev. Lett {\bf 98} 088702 

\bibitem{durand} 
M.~Durand, Phys. 
{\it Structure of Optimal Transport Networks Subject to a Global Constraint},
2007 Rev. Lett. {\bf 98} 088701

\bibitem{opper1999}
M. Opper and D. Saad, eds,
1999 {\it Advanced Mean Field Methods}
(Cambridge, MA: MIT Press)

\bibitem{vicente1999}
R. Vicente, D. Saad and Y. Kabashima,
{\it Finite-connectivity systems as error-correcting codes},
1999 Phys. Rev. E {\bf 60} 5352 

\bibitem{mackay2003}
D. J. C. Mackey, 
2003 {\it Information Theory, Inference and Learning Algorithms}
(UK: Cambridge University Press)

\bibitem{mezard2002}
M. M\'ezard and R. Zecchina,
{\it Random K-satisfiability problem: From an analytic solution to an efficient algorithm},
2002 Phys. Rev. E {\bf 66} 056126;
M. M\'ezard, G. Parisi and R. Zecchina, 
{\it Analytic and algorithmic solution of random satisfiability problems},
2002 Science {\bf 297} 812


\bibitem{jordan}
M. I. Jordan, ed., 
1999 {\it Learning in Graphical Models}
(Cambridge, MA: MIT Press)

\bibitem{mezard1987}
M. M\'ezard, G. Parisi and M. A. Virasoro
1987 {\it Spin Glass Theory and Beyond}
(World Scientific)

\bibitem{kelly1991}
F. P. Kelly,
{\it Network routing},
1991 Phil. Trans. R. Soc. Lond. A
{\bf 337} 343

\bibitem{barabasi}
A. L. Barab\'asi and R. Albert, 
{\it Emergence of Scaling in Random Networks},
1999 Science {\bf 286} 509

\end{thebibliography}
\end{document}